\documentclass[preprint,review,3p,twocolumn]{elsarticle}

\usepackage{amsmath,amsfonts}
\usepackage[dvipsnames]{xcolor}
\usepackage[ruled,vlined,linesnumbered,longend]{algorithm2e}

\usepackage{graphicx}
\usepackage{caption}
\usepackage{subcaption}
\usepackage{url}

\usepackage{comment}

\journal{Computer Communications}
    
\begin{document}

\begin{frontmatter}

\author{Theofanis P. Raptis}
\ead{theofanis.raptis@iit.cnr.it}
\author{Andrea Passarella}
\ead{andrea.passarella@iit.cnr.it}
\author{Marco Conti}
\ead{marco.conti@iit.cnr.it}

\address{Institute of Informatics and Telematics, National Research Council, Pisa, Italy}

%\title{Socially Aware Peer-to-Peer Wireless Energy Sharing Protocols}
%\title{Introducing Social Awareness in Peer-to-Peer Wireless Energy Sharing through Online Social Network Information}
\title{Energy Efficient Network Path Reconfiguration for Industrial Field Data\tnoteref{mytitlenote}}
\tnotetext[mytitlenote]{A preliminary version of this paper was presented in the 16th IFIP WG 6.2 International Conference on Wired/Wireless Internet Communications (WWIC 2018) \cite{10.1007/978-3-030-02931-9_3}}

\begin{abstract}
Energy efficiency and reliability are vital design requirements of recent industrial networking solutions. Increased energy consumption, poor data access rates and unpredictable end-to-end data access latencies are catastrophic when transferring high volumes of critical industrial data in strict temporal deadlines. These requirements might become impossible to meet later on, due to node failures, or excessive degradation of the performance of wireless links. In this paper, we focus on maintaining the network functionality required by the industrial, best effort, low-latency applications after such events, by sacrificing latency guarantees to improve energy consumption and reliability. We avoid continuously recomputing the network configuration centrally, by designing an energy efficient, local and distributed path reconfiguration method. Specifically, given the operational parameters required by the applications, our method locally reconfigures the data distribution paths, when a network node fails. Additionally, our method also regulates the return to an operational state of nodes that have been offline in the past. We compare the performance of our method through simulations to the performance of other state of the art protocols and we demonstrate performance gains in terms of energy consumption, data delivery success rate, and in some cases, end-to-end data access latency. We conclude by providing some emerging key insights which can lead to further performance improvements.
\end{abstract}


\end{frontmatter}

\section{Introduction}

%With the introduction of Internet of Things (IoT) concepts in industrial application scenarios, industrial automation is undergoing a tremendous change. This is made possible in part by recent advances in technology that allow interconnection on a wider and more fine-grained scale \cite{7883994}. The core of distributed automation systems and networks is essentially the reliable exchange of data. Any attempt to steer processes independently of continuous human interaction requires, in a very wide sense, the flow of data between some kind of sensors, controllers, and
%actuators \cite{6042563}.

The natural evolution of the data enabling industrial technologies and services leads to the generation of huge amounts of data; data of many different volumes, traffic and criticality. Data is serving as a fundamental resource to promote Industry 4.0 from machine automation to information automation and then to knowledge automation. Also, data will enable fast control cycles for applications such as zero-defect manufacturing, allowing information sharing across production sites of a given factory operator, or across value chains composed by different stakeholders \cite{8764545}.

Wireless industrial field deployments are becoming more prevalent in most industrial companies. Indeed, several industries have recently migrated from wired to wireless to take advantage of the cost benefits and infrastructure advancements \cite{8115313}. This type of networks ensures a wide range of industrial applications as it provides great advantages over traditional wired systems. Cost reduction can be achieved, in particular, by replacing the existing cables with a wireless medium, as long as an appropriate level of service for critical applications can still be guaranteed at all times \cite{8412519}. However, energy consumption remains the performance limiting factor and the biggest constraint for wireless industrial field. Currently, most industrial applications request battery life of several months and low power wireless is not viable in applications that require relatively large amounts of power \cite{6864830}.

In today's typical industrial field configurations, the computation of the data exchange and distribution schedules is quite primitive and highly centralized. Usually, the generated data are transferred to a central network controller or intermediate network proxy nodes using multi-radio industrial workcells which include high numbers of sensors and actuators communicating \cite{8694791}. In order to extend the network beyond the radio coverage of one node, a mesh topology enables a node to act as a relay for others, but, beyond one hop, the topology requires a protocol for routing data throughout the network, as well as novel data distribution algorithms \cite{8412519}. The controller analyzes the received information and, if needed, reconfigures the network paths and the data forwarding mechanisms, and changes the behavior of the physical environment through actuator devices. Traditional data distribution schemes can be implemented over relevant industrial protocols and standards, such as IEC WirelessHART, IEEE 802.15.4e and IETF 6TiSCH. Those entirely centralized and offline computations regarding data distribution scheduling, can become inefficient in terms of energy, when applied in industrial field deployments \cite{8793020}. In such environments, the topology and connectivity of the network may vary due to link and node failures \cite{7058728}. Furthermore, sensors may also be subject to RF interference, highly caustic or corrosive environments, high humidity levels, vibrations, dirt and dust, or other conditions that challenge performance \cite{us2002}. Also, very dynamic conditions, which make communication performance much different from when the central schedule was computed, possibly causing sub-optimal performance, may result in not guaranteeing application requirements. These dynamic network topologies may cause a portion of industrial sensor nodes to malfunction. With the increasing number of involved battery-powered devices, industrial field networks may consume substantial amounts of energy; more than needed if local, distributed computations were used instead. 

\textbf{Our contribution.} In this paper, we examine wireless industrial field networks comprised of sensing and actuating nodes, whereby data producers and consumers are known. We consider best effort, low-latency applications (that might require a certain maximum delivery delay from producers to consumers) for which at some point in time, a central controller computes a near-optimal set of multi-hop paths from producers to consumers, which guarantee a maximum delivery delay, while maximizing the energy lifetime of the networks (i.e., the time until the first node in the network exhaust energy resources). We focus on maintaining the network configuration in a way such that application requirements are considered even after important network operational parameters change due to some unplanned events (e.g., heavy interference, excessive energy consumption), while guaranteeing an appropriate use of nodes energy resources, by sacrificing latency guarantees to improve energy consumption. %In order to achieve this, we derive an energy efficient local and distributed path reconfiguration method. Specifically, 
We provide several efficient algorithmic functions which reconfigure the paths of the data distribution process, when a network node fails. %, or its conditions deteriorate too much with respect to the point in time when the original configuration was centrally determined. 
The functions regulate how the local path reconfiguration should be implemented and how a node can join a new path or modify an already existing path, ensuring that there will be no loops. Additionally, our functions also cover the cases when a node returns back to operational state, after past disactivations. The proposed method can be implemented on top of existing data forwarding schemes designed for industrial field networks. We demonstrate through simulations the performance gains of our method in terms of energy consumption and data delivery success rate, compared also to other state of the art solutions. % compared to other state of the art solutions.
  More specifically, the rest of this paper is organized as follows:
 \begin{itemize}
 \item In section \ref{sec::related}, we present the most related, recent works in the literature, focusing on industrial routing protocols.
 \item In section \ref{sec::back}, we provide the industrial field background, and we formulate the model used in this paper. 
 \item In section \ref{sec::epoch}, we formulate the \emph{epoch} a useful lifetime-based metric. This metric will assist the algorithmic design process.
 \item In section \ref{sec::pathrecon}, we present the algorithmic design of our method. The objective of our design is to provide a sustainable, energy efficient alternative to centralized data distribution solutions which dominate the industrial state of the art, in networks which come with node and link disactivations and reactivations on the initial set of nodes provides.
\item In section \ref{sec::pereval}, we present the performance evaluation results through simulation. We compare the performance of our method to other state of the art solutions in order to highlight the strengths and weaknesses of each.
\item In section \ref{sec::conc}, we conclude the paper and we provide some key directions for future reasearch.
 \end{itemize}

\section{Related work} \label{sec::related}

The combination of three elements of energy efficient path reconfiguration for industrial field data, namely, the industrial field communication (wireless, multi-radio, multi-hop), the industrial requirements (delay constraints, energy restrictions, and data pieces in need to be accessed by consumers), and the dynamic insertion and deletion of nodes in the network at any time, is closely related to industrial routing protocols. Proposed mechanisms for distributed path reconfiguration can act complementarily and on top of industrial routing protocols. For this reason, we provide a brief overview on the recent state of the art and on some classic industrial routing alternatives. The typical routing alternative for industrial low power deployments is IETF RPL \cite{rfc6550}. RPL aims to generate routing links between the network devices. Within RPL, the coordinating network controller has all the necessary information to incorporate different devices into the network. Each set of nodes within a network is part of one or more directed acyclic graphs. In each graph, once a root node is defined, the graph will be oriented to it, propagating the generated data towards the root node. In \cite{doi:10.1155/2014/936379} and \cite{7506102}, the authors present the reliable real-time flooding-based routing protocol REALFLOW. Contrary to the RPL case, instead of traditional routing tables, related node lists are generated in a distributed manner, serving for packet forwarding. A controlled flooding mechanism is applied to improve both reliability and real-time performance. The authors of \cite{8338161} propose a routing scheme that enhances energy consumption and end-to-end delay for large-scale industrial deployments based on IEEE 802.15.4a. The proposed scheme targets networks where data are aggregated through different clusters on their way to the sink. Moreover, a hierarchical system framework is employed to promote scalability, by estimating the residual energy and hop counts for each path. \cite{SEPULCRE2016121} contributes towards maintaining quality of service in industrial networks with a multi-path routing protocol that identifies and establishes the necessary redundant routes between any pair of nodes of a wireless network in order to satisfy the reliability and delay quality of service levels demanded by industrial applications.

\section{Industrial field background} \label{sec::back}

\begin{figure}[t!]
\centering
        \includegraphics[width=\columnwidth]{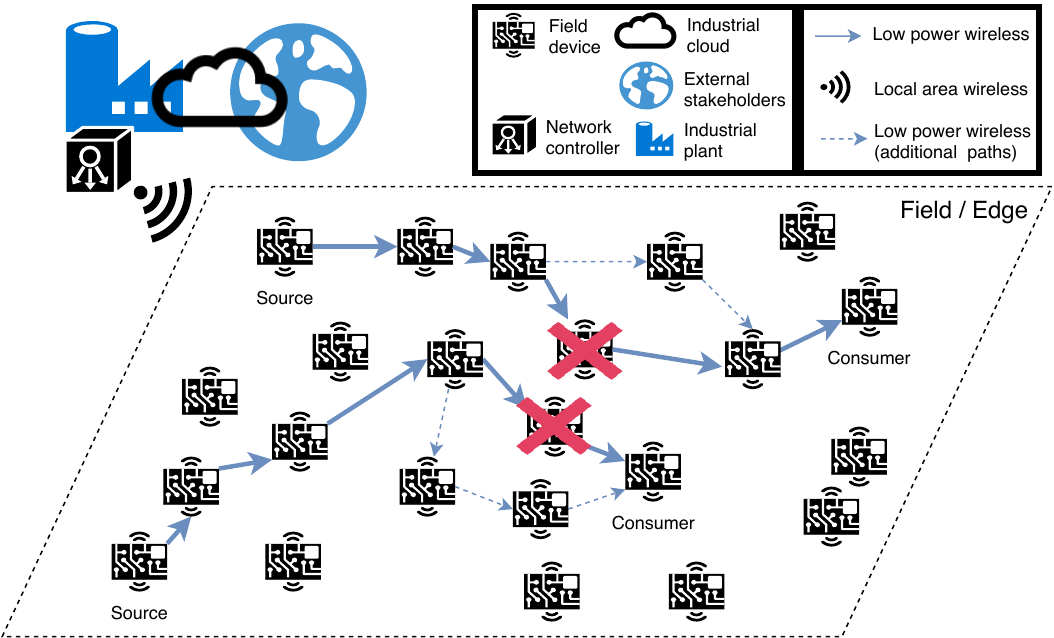}
        \caption{Industrial field architecture.}
        \label{fig::arch}
\end{figure}

In this section, we describe the technological background of the industrial field networks and we sketch the network model abstraction that we use in this paper. We consider industrial field deployments as shown in Figure~\ref{fig::arch}. The industrial plant is the center of operations and is connected to the industrial field on the one end and to the external stakeholders on the other end. The industrial plant features a network controller and the field contains the industrial network. The industrial network typically features a large number of sensor and actuator nodes (which are equipped with processing, storing and communication abilities). Such types of industrial networks are particularly related to smart factories \cite{8333734} and industrial workshop deployments \cite{8291116}. As shown in Figure~ \ref{fig::arch}, nodes (and wireless links) can go offline, and need to be substituted by other nodes and links.

Various communication and networking technologies co-exist in the industrial field deployment. Typically, the communication between the industrial plant and the external stakeholders is performed via an industrial cloud system and relevant services\footnote{The investigation of data distribution aspects on this part of the communications is outside the scope of this paper.}. The communication between the industrial plant and the field network is performed through the central network controller. The communication is usually implemented using two kinds of wireless technologies: local area radio for the inter-field communication between the field nodes and the controller (e.g., IEEE 802.11) and low power radio for the intra-field communication among the field nodes (e.g., IEEE 802.15.4). This multiple radio technology communication is particularly effective for use cases necessitating multiple radio technologies that are emerging in networked manufacturing shopfloor settings, especially within the frame of Industry 4.0 requirements. An example is the IK4-Tekniker use case\footnote{AUTOWARE collaborative workcell for assembly and logistics: \url{http://autoware-eu.org/case\_stories/collaborative\_workcell.php}} of the H2020 FoF AUTOWARE project \cite{Lucas_Esta__2018}, in which multiple communication technologies (TSN, IEEE 802.15.4 and IEEE 802.11g) co-exist in each industrial workcell node for redundancy and reliability. Another example is the H2020 FoF SatisFactory project\footnote{SatisFactory project: \url{http://www.satisfactory-project.eu/}.}, where a multiple radio smart factory use case is presented in \cite{LigiosDCRSP18}, in which nodes equipped with both 6LoWPAN and Wi-Fi interfaces, targeting energy efficiency and increased data rates.

We are particularly interested in the industrial field network of the field devices, as well as in the associated data generation. We model the industrial field network as a graph $G=(V,E)$, where $u \in V$ is a resource constrained sensor and actuator node and $E$ is the set of wireless links in the network. We assume that each wireless link is either fully enabled with perfect delivery reliability or fully disabled\footnote{The simulation results of our previous works \cite{raptisnew} and \cite{Raptis_2018} validate the fact that, for the considered scenario where the network is carefully deployed, this graph modeling assumption is a fairly good choice (the abstraction of a wireless network as a graph is debatable, mainly as the links between the nodes can be quite dynamic. This has been largely debated in community and as a result, stochastic geometric models have gained quite a bit of popularity. However, for the considered scenario where the network is carefully deployed this assumption is a fairly good one).}. We denote as $C$ the central network controller. We divide time $t$ in equal, discrete time intervals $\tau$. Every node $u \in V$, at time $t$, has an available amount of finite energy $E^t_u$ ($E^0_u$ at the beginning of the network operation). A node $u \in V$ can achieve one-hop data propagation using suitable industrial wireless technologies to a set of nodes which lie in its neighborhood $N_u$. $N_u$ contains the nodes $v \in V$ for which it holds that $\rho_u \geq \delta(u,v)$, where $\rho_u$ is the transmission range of node $u$ (defined by the output power of the antenna) and $\delta(u,v)$ is the Euclidean distance between $u$ and $v$. The sets $N_u$ are thus defining the set of edges $E$ (wireless links) of the graph $G$. Each one-hop data propagation from $u$ to $v$ results in a latency $l_{uv}$.  Assuming that all network nodes operate with the same output power, each one-hop data propagation at time $t$ from $u$ to $v$ requires an amount of $\epsilon_{uv}^t$ of energy dissipated by $u$ so as to transmit one data piece to $v$. A node can also transmit control messages to the network controller $C$ by consuming $\epsilon_{cc}^t$ amount of energy.  This wireless local area communication, however, is much more power hungry and thus more costly, as for example, low-power wireless links are typically operating at -25 dBm whereas local area wireless links at 15 dBm. So, for this kind of transmissions, we assume that more expensive wireless technology is needed, and thus we have that $\epsilon_{cc}^{t_1} \gg \epsilon_{uv}^{t_2}$, $\forall t_1, t_2$ (for example, the former can occur over WiFi or LTE links, while the latter over 802.15.4 links). 

%\subsection{The data}

The data generated in the network are modeled as a set of data pieces $D$. Each data piece $d \in D$ is defined as a triplet $d = (s_d, c_d, r_d)$, where $s_d \in V$ is the source of data piece $d$, $c_d \in V$ is the consumer\footnote{If the same data of a source, e.g., $s_1$, is requested by more than one consumers, e.g., $c_1$ and $c_2$, we have two distinct data pieces, $d_1 = (s_1, c_1, r_1)$ and $d_2 = (s_2, c_2, r_2)$, where $s_1 = s_2$.} of data piece $d$, and $r_d$ is the data generation rate of $d$. Each data piece $d$ is circulated in the network through a multi-hop path $\Pi_{d}$. Each node $u \in \Pi_{d}$ knows which is the previous node $previous(d,u) \in \Pi_{d}$ and the next node $next(d,u) \in \Pi_{d}$ in the path of data piece $d$.  We assume that the data is generated (according to rate $r_d$) at each source $s_d$ at the beginning of each time interval $\tau$, and circulated during the same time interval $\tau$. %The data generation and request patterns are not necessarily synchronous, and therefore, the data need to be cached temporarily for future requests by consumers. This asynchronous data distribution is usually implemented through an industrial pub/sub system \cite{8115846}. 
%\subsection{The latency constraint}
%Due to the fact that the set $P$ of proxy nodes is strong in terms of computation, storage and energy supplies, nodes $p \in P$ can act as proxy in the network and cache data originated from the sources, for access from the consumers when needed. This relieves the IoT devices from the burden of storing data they generate (which might require excessive local storage), and helps meeting the latency constraint. Proxy selection placement strategies have been studied in recent literature \cite{8115846,7876128}. 
 A critical aspect in the industrial operation is the timely data access by the consumers upon request, and, typically, the data distribution system must guarantee that a given maximum data access latency constraint (defined by the specific industrial process) is satisfied. We denote this threshold as $L_{\textnormal{max}}$. We denote as $L_{d}$ the latency of the multi-hop data propagation of the path $\Pi_{d}$, where $L_{d} = l_{uw} + ... + l_{zv}$. %When a data piece $d$ is generated at source $s_i$, it is delivered and stored to a proxy $p$ via a multi-hop wireless path. 
%Upon a request from $c_i$, data piece $d$ can be delivered from $p$ via a (distinct) multi-hop path. We denote as $L_{c_i}$ the data access latency of $c_i$, with $L_{c_i} = L_{c_ip} + L_{pc_i}$. 
 We assume an existing mechanism of initial centrally computed configuration of the data forwarding paths in the network, e.g., as presented in \cite{8390794}. In typical industrial settings, the requirements necessitate that the following constraint should be met: $L_{d} \leq L_{\textnormal{max}}, \forall d \in D$.

\section{Epochs and their maximum duration} \label{sec::epoch}
%\subsection{The epochs}

%In order to formulate the temporal aspects of the model we consider, we introduce the concept of epochs and we model the network as a time varying network \cite{2000klein,1337732}. 
In order to better formulate the data forwarding process through a lifetime-based metric, we define a time focused quantity, the \emph{epoch}. An epoch $j$ denotes the time $t_j$ elapsed between two consecutive, significant changes in the main network operational parameters ($t_j$ is a multiple of $\tau$), from the point of view of a single node. Specifically, when a node is about to get offline due to unexpected conditions, or when a node is changing its data distribution schedule, we say that this node changes epoch.
%A characteristic example of such change is a sharp increase of $\epsilon_{uv}^{\tau+1} \gg \epsilon_{uv}^\tau$ between two consecutive time intervals $\tau$ and $\tau+1$, due to sudden, increased interference on node $u$, which in turn leads to increased retransmissions on edge $(u,v)$ and thus higher energy consumption. In other words, $\frac{\epsilon_{uv}^{\tau} - \epsilon_{uv}^{\tau-1}}{\epsilon_{uv}^\tau} > \gamma$, where $\gamma$ is a predefined threshold. 
At the beginning of each epoch, (all or some of) the nodes initially take part in a configuration phase (centralized or distributed), during which they acquire the plan for the data distribution process by spending an amount of $e_u^{\textnormal{cfg}}$ energy for communication. Then, they run the data distribution process. An epoch is thus comprised of two phases: 
\begin{itemize}
\item \emph{Configuration phase.} During this initial phase, the nodes acquire the set of neighbors from/to which they must receive/forward data pieces in the next phase. 
\item \emph{Data forwarding phase.} During this phase the data pieces are circulated in the network according to the underlying network directives.
\end{itemize}

Epochs are just an abstraction that is useful for the design and presentation of the algorithmic functions, but does not need global synchronization. As it will be clear later on, when employing distributed solutions, each node locally identifies the condition for which an epoch is finished from its perspective, and acts accordingly (different nodes ``see'' in general different epochs). Although some events which affect the epoch duration cannot be predicted and thus controlled, we are interested in the events which could be affected by the data distribution process and which could potentially influence the maximum epoch duration. We observe that an epoch cannot last longer than the time that the next node in the network dies. Consequently, if we manage to maximize the time until a node dies due to energy consumption, we also make a step forward for the maximization of the epoch duration.

We now define the maximum epoch duration, due to the fact that it can serve as a useful metric for the decision making process of the distributed path reconfiguration. The maximum epoch duration is the time interval between two consecutive node deaths in the network. Specifically, each epoch's duration is bounded by the lifetime of the node with the shortest lifetime in the network, given a specific data forwarding configuration. Without loss of generality, we assume that the duration of the configuration phase equals $\tau$. We define the variables, $x^{dj}_{uv}$ which hold the necessary information regarding the transmission of the data pieces across the wireless links of the network. More specifically, for epoch $j$, $x^{dj}_{uv} = 1$ when the link $(u,v)$ is activated for data piece $d$. On the contrary, $x^{dj}_{uv}=0$ when the link $(u,v)$ is inactive for the transmission of data piece $d$. We denote as $a^j_{uv} = \sum_{d \in D} r_d x^{dj}_{uv}$ the aggregate data rate of $(u,v)$ for epoch $j$. Stacking all $a^j_{uv}$ together, we get $\mathbf{x}_u^j = [a^j_{uv}]$, the data rate vector of node $u$ for every $v \in N_u$, for epoch $j$. Following this formulation (and if we assume that $t_j \rightarrow \infty$) the maximum lifetime of $u$ during epoch $j$ can be defined as:
%\begin{equation}
%T_u(\mathbf{x}_j) = \frac{E_u}{\sum_{v \in N_u} \epsilon_{uv} a^j_{uv}}.
%\label{eq::nodelife}
%\end{equation}

\begin{equation} 
T_u(\mathbf{x}_u^j) =
  \begin{cases}
    \frac{E^j_u}{\sum_{v \in N_u} \epsilon_{uv} a^j_{uv}}       & \quad \text{if } E^j_u > e_u^{\textnormal{cfg}}\\
    \tau  & \quad \text{if } E^j_u \leq e_u^{\text{cfg}}\\
    0  & \quad \text{if } E^j_u = 0
  \end{cases},
  \label{eq::eq}
\end{equation}
where $e_u^{\textnormal{cfg}}$ is the amount of energy that is needed by $u$ in order to complete the configuration phase. Consequently, given an epoch $j$, the maximum epoch duration is $t_j^{\text{max}} = \min_{u \in V} \left\{ T_u(\mathbf{x}_u^j) \mid \sum_{v \in N_u} x^{dj}_{uv} > 0\right\}$. 

There have already been works in the literature which identify, for each data source $s_d$, %the proxy $p$ where its data should be cached, in order to maximise the total lifetime of the network until the first node dies \cite{draft} (or, in other words, 
how to maximize the time until the first node in the network dies (a time interval which coincides with the duration of the first epoch: $\max \min_{u \in V} \left\{ T_u(\mathbf{x}_u^1) \mid \sum_{v \in N_u} x^{d1}_{uv} > 0\right\}$ %), 
 and configure the data forwarding paths accordingly, while respecting the data access latency constraints \cite{8390794}. Reconfigurations can be triggered also when the conditions under which a configuration has been computed, change. Therefore, (i) epoch duration can be shorter than $t_j$, and (ii) we do not need any centralized synchronization in order to define the actual epoch duration. We consider the epoch as only an abstraction (but not a working parameter for the functions), which is defined as the time between two consecutive reconfigurations of the network, following the functions presented in section \ref{sec::pathrecon}.

%However, due to the fact that industrial IoT networks have to operate beyond this time point, regardless of some nodes failures, the initial configuration necessitates online reconfigurations on the data forwarding paths when different nodes die. %The aim of this paper is to ensure that we are able to %$\max \min_{u \in V} \left\{ T_u(\mathbf{x}_j) \mid \sum_{v \in N_u} x^{ij}_{uv} > 0\right\}$, $\forall j$, 
% efficiently handle the network reconfigurations.

\section{Distributed path reconfiguration and data forwarding} \label{sec::pathrecon}

The objective of our design is to provide a sustainable, energy efficient alternative to centralized data distribution solutions which dominate the industrial state of the art, in networks which come with node disactivations and reactivations on the initial set of nodes provides. The intuition behind the expected impact of this decentralization is depicted in Table \ref{tab::trade}. Data distribution solutions such as the one presented in \cite{8390794}, can achieve a near optimal epoch duration, due to the precision of the global knowledge available for the configuration, as well as guaranteed compliance to the data access latency constraints. However, because the central reconfiguration process requires all nodes to communicate their status to the central network controller via the expensive local area links (instead of cheap low power), those solutions are expected to have more expensive reconfiguration phases of a much larger scale as well.

\begin{table}[t!]
\centering
\caption{Trade-offs.}
\begin{tabular}{ r||c| c }
& Centralized  \cite{8390794} & Distributed\\\hline
cfg. cost&$ \color{red} \mathcal{O}(\epsilon_{cc}^t \cdot |V|)$  &\color{ForestGreen} $\mathcal{O}(\epsilon_{uv}^t \cdot|D|\cdot|N_u|)$\\
%\textbf{data relocation cost}& \color{red} $\mathcal{O}(\epsilon_{uv}kDE)$&$ \color{ForestGreen} 0$\\
\textbf{$T_u$}&{\color{ForestGreen}near optimal} &\color{red} not optimal\\
\textbf{$L_{d} \leq L_{\textnormal{max}}$}&{\color{ForestGreen}guaranteed}& \color{red} not guaranteed
%\textbf{Parameter} & \textbf{Value}\\\hline
\\\hline
\end{tabular}
\label{tab::trade} 
\end{table}

The main idea behind our solution is the following: the nodes are initially provided with a centralized data forwarding plan. When a significant change in the network occurs, the nodes involved are locally adjusting the paths, using the lightweight communication links among them (e.g., 802.15.4) instead of communicating with the central network controller (e.g., LTE, WiFi). The main metric used for the path adjustment is the epoch-related $T_u(\mathbf{x}_u^j)$ lifetime, as defined in Eq.~\ref{eq::eq}.

The functions' pseudocode is presented in the following subsections. %Due to lack of space we omit the presentation of some functions' pseudocodes, but those can be found in the extended version of the paper \cite{extended}. 
The functions are presented in upright typewriter font and the messages which are being sent and received are presented in italics typewriter font. The arguments in the parentheses of the functions are the necessary information that the functions need in order to compute the desired output. The arguments in the brackets of the messages are the destination nodes of the messages and the arguments in the parentheses of the messages are the information carried by the messages. We now present the way that our solution handles the path reconfiguration distributively in the case when nodes go offline and in the case when nodes are returning online.

\subsection{Nodes going offline} \label{sec::off}

%When a significant change in the network happens due to interference, malfunction, or low energy supplies, some of the data distribution paths may get disconnected, or some nodes might start consuming much higher energy than expected. In this section we present the distributed path reconfiguration and data forwarding method. 

 We assume that a node $u$ complies with the following rules: $u$ knows the positions of every $v \in N_u$, $u$ knows the neighborhood $N_v$ of every node $v$ in its own neighborhood $N_u$, and $u$ stores only local information or temporarily present data pieces in transit. The acquisition from $u$ of the knowledge of $N_v$ requires additional energy costs only once, in the beginning of the network operation (as the transmission ranges are fixed and the neighborhoods do not change). Therefore, there is no need to keep consuming energy for maintaining or keeping updated this information. This is because the fact that a node might go offline or deplete its energy supplies, does not alter the neighborhood modeling structure.

\begin{algorithm*}[t!]
\DontPrintSemicolon
%\SetKwInOut{Input}{Input}\SetKwInOut{Output}{Output}
%$\forall u \in V$ with $E_u > 0$: $u$ communicates with the central controller\;
%\If{$u \textnormal{ is online}$}{ \label{algo::DistrDataFwd::init1}
\If{$t = 0$}{
%\tcc{nodes communicate their status to $C$}
send \texttt{\emph{status}}[$C$]$(E_u, \epsilon_{uv}^0, l_{uv}), \forall v \in N_u$\; \label{algo::DistrDataFwd::init1}
%\tcc{$C$ computes paths and data storage locations (e.g., \cite{draft})}
receive \texttt{\emph{plan}}[$u$]$(y^{d1}_{uv})$\;%\tcc{$\tau=1,j=1$}
%$\tau = 1$\; \label{algo::DistrDataFwd::init2}
$x^{d1}_{uv} = y^{d1}_{uv}, \forall d \in D, \forall v \in N_u$\; \label{algo::DistrDataFwd::init2}
}
\Else{
call \texttt{Revive}$(u)$\;
}
%\tcc{$u$ repeats until out of energy or until extreme interference}
%\Repeat{$E_u = 0$ or $\frac{\epsilon_{uv}(\tau) - \epsilon_{uv}(\tau-1)}{\epsilon_{uv}(\tau)} > \gamma$ for $>50\%$ of active edges $(u,v)$ of $u$}{ \label{algo::DistrDataFwd::rep1}
\Repeat{$u \textnormal{ is about to go offline}$}{ \label{algo::DistrDataFwd::rep1}
	run \texttt{DataForwarding}$(\tau)$\; \label{algo::DistrDataFwd::fwd}
%	\tcc{Deactivate single edge}
	%\If{$\exists (u,v)$ with $\frac{\epsilon_{uv}(\tau) - \epsilon_{uv}(\tau-1)}{\epsilon_{uv}(\tau)} > \gamma$}{ \label{algo::DistrDataFwd::msgs1} \label{algo::DistrDataFwd::de1}
	\If{$\exists (u,v) \textnormal{ with abnormal energy consumption}$}{ \label{algo::DistrDataFwd::msgs1} \label{algo::DistrDataFwd::de1}
	$x^{dj}_{uv} = 0$, $\forall d \in D$\;
	send \texttt{\emph{alert}}[$previous(d,u)$]$(d, u, v)$, $\forall d \in D$\;
	} \label{algo::DistrDataFwd::de2}
%	\tcc{Reconfigure path after alert reception}
	\If {receive \texttt{alert}[$u$]$(d, v, next(d, v))$}{ \label{algo::DistrDataFwd::lpc}
		$x^{dj}_{uv} = 0$\;
		call \texttt{LocalPathConfig}$(d,u,next(d,v))$\;%\tcc{$\tau++,j++$}		
	}
%	\tcc{Join path}
	\If {receive \texttt{join}[$u$]$(d,w,v)$}{ \label{algo::DistrDataFwd::joi}
		call \texttt{JoinPath}$(d,w,v)$\;%\tcc{$\tau++,j++$}		
	}
%	\tcc{Modify existing path and avoid loops}
	\If{receive \texttt{modify\_path}$[u](d,w,deleteArg,dirArg)$}{ \label{algo::DistrDataFwd::mod}
		call \texttt{ModifyPath}$(d,w,deleteArg,dirArg)$\;
	} \label{algo::DistrDataFwd::msgs2}
	\If{receive \texttt{info\_request}$[u](v)$}{ \label{algo::DistrDataFwd::mod1}
		send \texttt{\emph{info}}[$v$]$(x^{dj}_{uv}, T_u)$\; \label{algo::DistrDataFwd::mod2}
	\If{receive \texttt{\emph{plan}}$[u](y^{dj}_{uv})$}{ \label{algo::DistrDataFwd::mod3}
		$x^{dj}_{uv} = y^{dj}_{uv}, \forall d \in D, \forall v \in N_u$\;\label{algo::DistrDataFwd::mod4} 
		}
	} \label{algo::DistrDataFwd::msgs2}
$t++$\;
}\label{algo::DistrDataFwd::rep2}
%}
%\tcc{Send alert and disconnect}
send \texttt{\emph{alert}}[$previous(d,u)$]$(d, u, v)$, $\forall d \in D$, $\forall v \in N_u$\; \label{algo::DistrDataFwd::ciao}
\texttt{Disconnect}($u$)\;
\caption{\texttt{DistrDataFwd$(u)$}}
\label{algo::DistrDataFwd}
\end{algorithm*}

\textbf{Distributed data forwarding.} The distributed data forwarding function \texttt{DistrDataFwd}$(u)$ is the main function that  is being ran on every node $u$ of the network. Its pseudocode is provided in the body of Algorithm~\ref{algo::DistrDataFwd}. At first, when node $u$ goes online, it checks if $t = 0$ or if it reconnects after a previous disconnection. If $t = 0$, the node communicates its status to the central network controller (which uses the method presented in \cite{8390794} for computing the data distribution parameters in an initial setup phase of the network), it receives the data forwarding plan, which is expressed by the values $y$, and it configures the wireless data distribution links accordingly, by setting the variables $x$ equal to the received values $y$. (lines \ref{algo::DistrDataFwd::init1}-\ref{algo::DistrDataFwd::init2}). On the other hand, if $t > 0$, then the node has returned back online from a previous disconnection. In this case, $u$ runs the function \texttt{Revive}$(u)$, which provides $u$ with an updated data distribution plan derived locally from its neighbors (function \texttt{Revive}$(u)$ is explained in section \ref{sec::return} which is focusing on nodes returning online). Then, for every time interval $u$ repeats the following process, until %either it is almost dead, or more than half of its associated wireless links spend more energy compared to the previous time interval, according to the system parameter $\gamma$ 
 a condition of going offline is satisfied (lines \ref{algo::DistrDataFwd::rep1}-\ref{algo::DistrDataFwd::rep2}): $u$ starts the data forwarding process according to the data distribution plan received by $C$ (line \ref{algo::DistrDataFwd::fwd}). Afterwards, it checks if a set of control messages have been received from any $v \in N_u$ and acts accordingly, by calling the necessary functions (lines \ref{algo::DistrDataFwd::msgs1}-\ref{algo::DistrDataFwd::msgs2}).

\begin{algorithm*}[t!]
\DontPrintSemicolon
%\SetKwInOut{Input}{Input}\SetKwInOut{Output}{Output}
%\While {$ E_u > 0$ and $\epsilon_{uv} < \epsilon_{\text{max}}$}{

%	\For{$i=1:d$}{
%		run data distribution\;
%		\tcc{Node $v$ stopped operating: incr. interference/low batt.}
%		\If{$v \in N_u \cap \Pi_{d}$ and $previous(d,v) = u$}{
%		\tcc{Replace $v$ with best lifetime $w$ and send join message to $w$} 
			\uIf{$\exists \iota \in N_u$ with $v \in N_{\iota}$ and $l_{u\iota} + l_{\iota v} \leq l_{uprevious(d,v)} + l_{previous(d,v)v}$}{ \label{algo::LocalPathConfig::chk}
				$w = \arg\max_{\iota \in N_u} T_{\iota}(\mathbf{x}_\iota^{j_\iota})$\; \label{algo::LocalPathConfig::rpl}
				send \texttt{\emph{join}}[$w$]$(d,u,v)$\; \label{algo::LocalPathConfig::send}
				$\Pi_{d} \leftarrow$ replace $v$ with $w$\;
				%\If{receive \texttt{\emph{modify\_path}}$[u](d,w,joinArg,deleteArg,dirArg)$}{
				%	\tcc{If no loop exists or forward loop exists}
				%	\uIf{$joinArg = joinYES$ and $deleteArg = deleteNO$}{
				% 		$\Pi_{d} \leftarrow$ replace $v$ with $w$\;
				%	}
				%	\tcc{If backward loop exists}
				%	\ElseIf{$joinArg = joinYES$ and $deleteArg = deleteYES$}{
			%
				%	}
				%}
			}
%			\tcc{Explore extended neighbourhood and replace $v$ with $w,w'$}
			\Else{
				run \texttt{local\_aodv+}$(u,v,TTL)$ \label{algo::LocalPathConfig::aodv}
				
				}
%		}
		%\If {\texttt{\emph{received\_update}}$(v', next(d, v))$}{
		
		%}
%	}
%}
%\texttt{\emph{send\_alert}}$(u, next(d, u))$
\caption{\texttt{LocalPathConfig}$(d,u,v)$}
\label{algo::LocalPathConfig}
\end{algorithm*}

If $u$ detects that a link is consuming too much energy and has to be deactivated, it deactivates this link (by causing a path disconnection) and notifies the previous node in the path for $d$, $previous(d,u)$, by sending an alert message (lines \ref{algo::DistrDataFwd::de1}-\ref{algo::DistrDataFwd::de2}). For a given deactivated link $(u,v)$ for data piece $d$, alert messages contain information on which is the data piece of interest, and which were the two nodes $u,v$ in the path prior to disconnection. Then, $u$ checks whether there has been an alert message received (line \ref{algo::DistrDataFwd::lpc}), and calls function \texttt{LocalPathConfig} (displayed in Algorithm~\ref{algo::LocalPathConfig}). Through this function the paths can be reconfigured accordingly, for all involved data pieces $d$. Due to the fact that \texttt{LocalPathConfig} sends some additional messages regarding joining a new path and modifying an existing one, $u$ then checks for reception of any of those messages (lines \ref{algo::DistrDataFwd::joi} and \ref{algo::DistrDataFwd::mod}) and calls the necessary functions \texttt{JoinPath} and \texttt{ModifyPath}. Lines \ref{algo::DistrDataFwd::mod1}-\ref{algo::DistrDataFwd::mod4} refer to the process followed after receiving an information request by a revived node and they are explained in section \ref{sec::return}. Finally, $u$ sends an alert message to the previous nodes in the existing paths prior to final disconnection due to energy supplies shortage (line \ref{algo::DistrDataFwd::ciao}).

\textbf{Local path configuration.} A node $u$ calls the path configuration function \texttt{LocalPathConfig} when it receives an alert which signifies cease of operation of a wireless link $(u,v)$ due to a sudden significant increase of energy consumption due to interference %$\left(\frac{\epsilon_{uv}(\tau) - \epsilon_{uv}(\tau-1)}{\epsilon_{uv}(\tau)} > \gamma \right)$ 
 or a cease of operation of a node $v$ due to heavy interference in all of $v$'s links or due to low energy supplies (Algorithm~\ref{algo::DistrDataFwd}, lines \ref{algo::DistrDataFwd::de2} and \ref{algo::DistrDataFwd::ciao}). %\texttt{LocalPathConfig}$(u,v)$ is also called when $u$ detects that the link $(u,v)$ is draining too much energy. 

 \texttt{LocalPathConfig} is inherently local and distributed. The goal of this function is to restore a functional path between nodes $u$ and $v$ by replacing the problematic node $previous(d,v)$ with a better performing node $w$, or if $w$ does not exist, with a new efficient multi-hop path $\Pi_{d}$. At first, $u$ checks if there are nodes $\iota$ in its neighborhood $N_u$ which can directly connect to $v$ and achieve a similar or better one-hop latency than the old configuration (line \ref{algo::LocalPathConfig::chk}). If there are, then the $w$ selected is the node $\iota$ which given the new data piece, will achieve a maximum lifetime compared to the rest of the possible replacements, i.e., $w = \arg\max_{\iota \in N_u} T_\iota(\mathbf{x}_u^{j_\iota})$, and an acceptable latency $l_{uw} + l_{wv}$ (line \ref{algo::LocalPathConfig::rpl}). $u$ then sends to $w$ a \texttt{\emph{join}} message (line \ref{algo::LocalPathConfig::send}).
 
If such a node does not exist, then $u$ runs \texttt{local\_aodv+}, a protocol that we introduce in this paper. We design \texttt{local\_aodv+} as a slightly modified, local version of the well-known AODV protocol for route discovery, between nodes $u$ and $v$. \texttt{local\_aodv+} is able to add more than one replacement nodes in the path. The main modification of \texttt{local\_aodv+} with respect to the traditional AODV protocol is that \texttt{local\_aodv+} selects the route which provides the  maximum lifetime $T_w(\mathbf{x}_u^j)$ for the nodes $w$ which are included in the route. Specifically, this modification with respect to the classic AODV is implemented as follows: The nodes piggyback in the route request messages the minimum lifetime $T_w(\mathbf{x}_u^j)$ that has been identified so far on the specific path. Then when the first route request message arrives at $v$, instead of setting this path as the new path, $v$ waits for a predefined timeout for more route request messages to arrive. Then, $v$ selects the path which provided the $\max\min_{w \in N_u} T_w(\mathbf{x}_u^{j_w})$. We use time to live (TTL) or hop limit as the mechanism which limits the lifespan or route discovery messages in the network. The propagation distance can be restrained by setting the TTL field, in order to prevent unnecessary network-wide dissemination, as well as undesirably high energy consumption rates. The reader can find more details about the AODV protocol in \cite{aodv}.

\begin{algorithm}[t!]
\DontPrintSemicolon
%\tcc{If $u$ is not in the path already, there is no loop.}
\uIf{$u \notin \Pi_{d}$}{
$previous(d,u) = w$\;
$next(d,u) = v$\;
send \texttt{\emph{modify\_path}}[$v$]$(d,u,deleteNO,fwd)$\;
%send \texttt{\emph{modify\_path}}[$w$]$(d,u,joinYES,deleteNO,null)$\;
}
%\tcc{Forward loop: $u$ is in the path already, after $w$.}
\uElseIf{$u \in \Pi_{d}$ and $u>w$}{
$previous(d,u) = w$\;
%send \texttt{\emph{modify\_path}}[$w$]$(d,u,deleteNO,null)$\;
send \texttt{\emph{modify\_path}}[$v$]$(d,u,deleteYES,fwd)$\;
}
%\tcc{Backward loop: $u$ is in the path already, before $w$.}
\ElseIf{$u \in \Pi_{d}$ and $u<w$}{
$next(d,u) = v$\;
send \texttt{\emph{modify\_path}}[$v$]$(d,u,deleteNO,fwd)$\;
send \texttt{\emph{modify\_path}}[$w$]$(d,u,deleteYES,bwd)$\;
}
\caption{\texttt{JoinPath}$(d,w,v)$}
\label{algo::JoinPath}
\end{algorithm}

\begin{algorithm*}[t!]
\DontPrintSemicolon
%		\tcc{Modification with edge deletions}
		\uIf{$deleteArg = deleteYES$ and $u \neq w$}{
%		\tcc{Forward modification}
			\uIf{$dirArg = fwd$}{
				send \texttt{\emph{modify\_path}}[$next(d, u)$]$(d,w,deleteYES,fwd)$\;
				\texttt{Deactivate}$(d,(u,next(d, u)))$\;
			}
%		\tcc{Backward modification}
			\ElseIf{$dirArg = bwd$}{
				send \texttt{\emph{modify\_path}}[$previous(d, u)$]$(d,w,deleteYES,bwd)$\;
				\texttt{Deactivate}$(d,(u,previous(d, u)))$\;			
			}
		}
%		\tcc{Modification without edge deletions}
		\ElseIf{$deleteArg = deleteNO$}{
%		\tcc{Forward modification}
			\uIf{$dirArg = fwd$}{
				%send \texttt{\emph{modify\_path}}[$next(d, u)$]$(d,w,deleteYES,fwd)$\;
				%\texttt{Deactivate}$(d,(u,next(d, u)))$\;
				$previous(d,u) = w$\;
			}
%		\tcc{Backward modification}
			\ElseIf{$dirArg = bwd$}{
				%send \texttt{\emph{modify\_path}}[$previous(d, u)$]$(d,w,deleteYES,bwd)$\;
				%\texttt{Deactivate}$(d,(u,previous(d, u)))$\;
				$next(d,u) = w$\;					
			}
		
		}
\caption{\texttt{ModifyPath}$(d,w,deleteArg,dirArg)$}
\label{algo::ModifyPath}
\end{algorithm*}

\textbf{Joining new paths, modifying existing paths and avoiding loops.} In this subsection, we briefly describe the functions regarding joining a new path and modifying an already existing path for loop elimination. %Due to lack of space, we do not include the pseudocode of those functions; however, they can be found at the extended version of this paper \cite{extended}. 
\texttt{JoinPath}$(d,w,v)$ is the function which regulates how, for data piece $d$, a node $u$ will join an existing path between nodes $w$ and $v$ and how $u$ will trigger a path modification and avoid potential loops which could result in unnecessary traffic in the network. Due to the fact that the reconfigurations do not use global knowledge, we can have three cases of $u$ joining a path: (i) $u$ is not already included in the path ($u \notin \Pi_{d}$), (ii) $u$ is already included in the path ($u \in \Pi_{d}$), and $w$ is preceding $u$ in the new path ($previous(d,u) = w$) with a new link $(w,u)$, and (iii) $u$ is already included in the path ($u \in \Pi_{d}$), and $u$ is preceding $w$ in the new path ($previous(d,w) = u$) with a new link $(u,w)$. In all three cases, \texttt{JoinPath} sends a modification message to the next node to join the path, with the appropriate arguments concerning the deletion of parts of the paths, and the direction of the deletion, for avoidance of potential loops. This messages triggers the function \texttt{ModifyPath}. In case (i) it is apparent that there is no danger of loop creation, so there is no argument for deleting parts of the path.

\begin{figure*}[t!]
\centering
    \begin{subfigure}[b]{0.32\textwidth}
    \centering
        \includegraphics[width=\columnwidth]{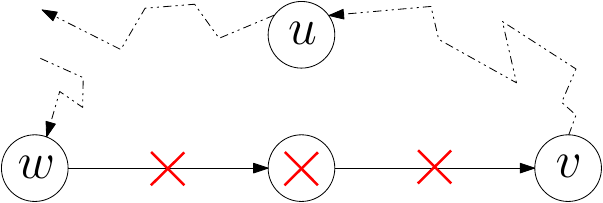}
        \caption{Unplanned change.}
        \label{fig::fwdloop1}
    \end{subfigure}
    \begin{subfigure}[b]{0.32\textwidth}
    \centering
        \includegraphics[width=\columnwidth]{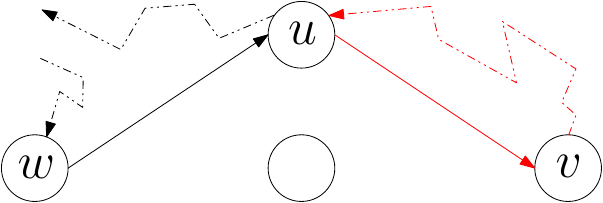}
        \caption{$u$ (re-)joining the path.}
        \label{fig::fwdloop2}
    \end{subfigure}   
    \begin{subfigure}[b]{0.32\textwidth}
    \centering
        \includegraphics[width=\columnwidth]{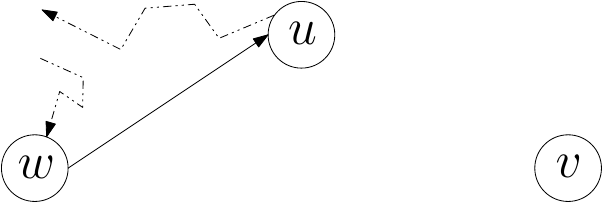}
        \caption{Loop elimination.}
        \label{fig::fwdloop3}
    \end{subfigure}   
\caption{Loop avoidance - forward loop.}
\label{fig::fwdloop}
\end{figure*}

\begin{figure*}[t!]
\centering
    \begin{subfigure}[b]{0.32\textwidth}
    \centering
        \includegraphics[width=\columnwidth]{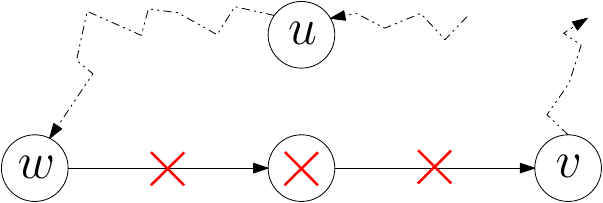}
        \caption{Unplanned change.}
        \label{fig::bwdloop1}
    \end{subfigure}
    \begin{subfigure}[b]{0.32\textwidth}
    \centering
        \includegraphics[width=\columnwidth]{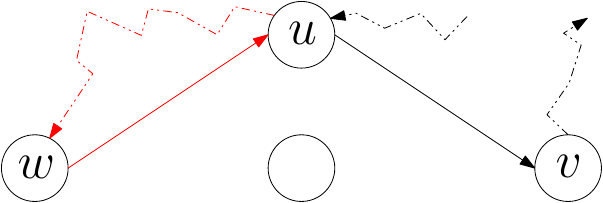}
        \caption{$u$ (re-)joining the path.}
        \label{fig::bwdloop2}
    \end{subfigure}   
    \begin{subfigure}[b]{0.32\textwidth}
    \centering
        \includegraphics[width=\columnwidth]{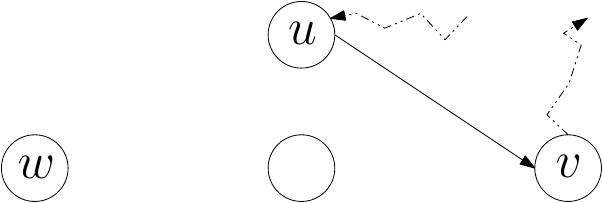}
        \caption{Loop elimination.}
        \label{fig::bwdloop3}
    \end{subfigure}   
\caption{Loop avoidance - backward loop.}
\label{fig::bwdloop}
%    \vspace{-0.7cm}
\end{figure*}

In order to better understand cases (ii) and (iii) we provide Figures \ref{fig::fwdloop} and \ref{fig::bwdloop}. In those Figures we can see how the function \texttt{ModifyPath} eliminates newly created loops on $u$ from path reconfigurations which follow unplanned network changes. In case (ii), which is depicted in Figure~\ref{fig::fwdloop}, $u$ and $w$ are both already in the path, $u$ after $w$ (Figure~\ref{fig::fwdloop1}) After an unplanned change in the network, a new path is connecting them for the second time (Figure~\ref{fig::fwdloop2}). Then \texttt{JoinPath}$(d,w,v)$ triggers function \texttt{ModifyPath} with arguments $deleteYES$ and $fwd$ so as to delete the existing loop between $u$ and $v$ (Figure~\ref{fig::fwdloop3}). In case (iii), which is depicted in Figure~\ref{fig::bwdloop}, $u$ and $w$ are both already in the path, $u$ before $w$ (Figure~\ref{fig::bwdloop1}). After an unplanned change in the network, a new path is connecting them for the second time (Figure~\ref{fig::bwdloop2}). Then \texttt{JoinPath}$(d,w,v)$ triggers function \texttt{ModifyPath} with arguments $deleteYES$ and $bwd$ so as to delete the existing loop between $u$ and $v$ (Figure~\ref{fig::bwdloop3}, red $\times$ marks). 

%\begin{comment}

%\end{comment}

Following the loop elimination process, loop freedom is guaranteed for the cases where there are available nodes $w \in N_u$ which can directly replace $v$. In the case where this is not true and \texttt{LocalPathConfig} calls \texttt{local\_aodv+} instead (Algorithm~\ref{algo::LocalPathConfig}, line \ref{algo::LocalPathConfig::aodv}), then the loop freedom is guaranteed by the AODV path configuration process, which has been shown to be loop free \cite{5371479}.

\subsection{Nodes returning online} \label{sec::return}

When a node $u \in V$, which has been previously offline due to unexpected event, returns again online with full operational capacity, the network reconfigures the data distribution according to the function \texttt{Revive}$(u)$, as follows. At first, $u$ sends a message to every node $w \in N_u$ so as to inform them about the return, and receives their response with the information on their data distribution links and remaining lifetime (lines \ref{revive::sendInfo}-\ref{revive::receiveInfo}). Note that, on the side of $w$, this process is being executed at Algorithm \ref{algo::DistrDataFwd}, lines \ref{algo::DistrDataFwd::mod1}-\ref{algo::DistrDataFwd::mod2}, where $w$ receives the information request and sends the related information. Afterwards, $u$ selects serially (at random) neighboring nodes and checks whether their remaining lifetime is lower than it own lifetime (line \ref{revive::lifetime}). If this is the case, $u$ replaces $w$ in the path of every data piece that is being served by $w$, locally recomputes the lifetimes before it proceeds to the next iteration and locally updates the new data distribution schedules storing them in the local variables $y^{dj_w}_{wv}$. When this iterative process ends, $u$ communicates the new schedules $y^{dj_w}_{wv}$ to the interested nodes (lines \ref{revive::communi}-\ref{revive::end}). When a node receives this information, it is changing the current wireless data distribution links accordingly (Algorithm \ref{algo::DistrDataFwd}, lines \ref{algo::DistrDataFwd::mod3}-\ref{algo::DistrDataFwd::mod4}).

\begin{algorithm*}[t!]
\DontPrintSemicolon
send \texttt{info\_request}$[w](u)$, $\forall w \in N_u$\; \label{revive::sendInfo}
receive \texttt{\emph{info}}[$u$]$(x^{dj_w}_{wv}, T_w)$\; \label{revive::receiveInfo}
\ForEach{$w \in N_u$}{
\If{$T_u > T_w$}{ \label{revive::lifetime}
\ForEach{$d \textnormal{ with } w \in \Pi_d$}{
\If{$previous(d,w), next(d,w) \in N_u$}{
call \texttt{JoinPath}$(d,previous(d,w), next(d,w))$\;
$\Pi_{d} \leftarrow$ replace $w$ with $u$\;
%update $x^{dj}_{uw}, x^{dj}_{vw}, x^{dj}_{previous(d,v)w}, x^{dj}_{next(d,v)w}$\;
update $T_u, T_w, T_{previous(d,w)}, T_{next(d,w)}$
}
$y^{dj_w}_{wv} = 1 \Leftrightarrow (w,v) \in \Pi_d, \forall v \in N_w$\;
}
}
}
\ForEach{$w \in N_u$}{ \label{revive::communi}
\If{$y^{dj_w}_{wv} \neq x^{dj_w}_{wv}$}{
send  \texttt{\emph{plan}}$[v](y^{dj_w}_{wv})$ \label{revive::end}
}
}
\caption{\texttt{Revive}$(u)$}
\label{algo::Revive}
\end{algorithm*}

\section{Performance evaluation} \label{sec::pereval}

%\begin{comment}

%\end{comment}

In order to implement and simulate \texttt{DistrDataFwd} we opted for Matlab. The main reasons for selecting Matlab are: (i) the ease of designing the real-time, dynamic multi-hop path reconfigurations, (ii) the abstract modeling options offered, which better reflect the paper's modeling approach and take away the burden of implementing the detailed networking aspects (PHY, MAC, etc.), (iii) the convenience of having ready and using as a black box, already implemented functions for designing the initial network path computation ($k$-shortest paths, etc.). Additionally, we used some results obtained from real low-power devices, characteristic for the type of applications that we consider, in order to set the latency parameters, as explained in the following.

\begin{table*}[t!]
\centering
\caption{Simulation parameters.}
\begin{tabular}{ l | l }\hline
\textbf{Parameter} & \textbf{Value}\\\hline
\multicolumn{2}{c}{\textbf{Topology}} \\\hline
deployment dimensions (2D grid) & $7.5$ m $\times$ $16.0$ m\\
$|V|$, $|E|$,  $\delta (u,v)$ & $18, 47$, $2.5 - 2.8$\\
transmission range $\rho_u$, neighborhood $N_u$ & $3$ m, $v$, with $\delta(u,v) \leq 3$ m\\\hline
\multicolumn{2}{c}{\textbf{Time}} \\\hline
$\tau$, $L_{\text{max}}$, $TTL$ (\texttt{local\_aodv+}) & $1$ sec, $100$ ms, $2$ hops \\
experiment duration & $2000$ hours\\\hline
\multicolumn{2}{c}{\textbf{Data}}\\\hline
percentage of consumers $c_d$ & $0.05 - 45\%$\\
data piece generation rate $r_d$ & $1-8$ $d/\tau$\\
data piece size (incl. headers and CRC) & $9$ bytes\\\hline
\multicolumn{2}{c}{\textbf{Hardware assumptions}} \\\hline
MCU (e.g., MSP430), antenna (e.g., CC2420) & ultra low-power, IEEE 802.15.4  \\
max. battery capacity, initial energies $E_u^0$ & $830$ mAh / $3.7$ V, $0-3$ Wh\\
transmission power for $e^t_{uv}$, for $e^t_{cc}$ & $-25$ dBm, $15$ dBm\\\hline
\end{tabular}
\label{tab::parameters}
\end{table*}

We configured the simulation environment according to realistic parameters and assumptions. The parameter configuration in details can be found in Table \ref{tab::parameters}. Briefly, we assume an industrial field network, comprised of devices equipped with ultra low-power MCUs, such as MSP430, and IEEE 802.15.4 antennae, such as CC2420, able to support industrial IoT standards and protocols, such as WirelessHART and IEEE 802.15.4e\footnote{In the current work we assume a simple CSMA scheme. In the case that a protocol designer wishes to align our methods to specific industrial wireless standard, additional design elements should be taken care of. For example, in the case of WirelessHART, as noted in \cite{8793020}, the channel scheduling mechanism has to be adapted according to the application requirements. Those design issues are left for future work and more details are provided in section \ref{sec::conc}.}. We assume a structured topology (as in usual controlled industrial settings) of 18 nodes which form a 2D grid with dimensions of $7.5$ m $\times$ $16.0$ m. We set the transmission power of the nodes for multi-hop communication to $-25$ dBm (typical low-power) which results in a transmission range of $3$ m. For the more expensive communication with the network controller, we set the transmission power to $15$ dBm, typical of wireless LAN settings. We set the time interval $\tau = 1$ second, the percentage of consumers over the nodes $0.05-45$\% and we produce $1-8$ $d/\tau$ per consumer. In order to perform the simulations in the most realistic way, we align the $L_{\text{max}}$ value with the official requirements of future network-based communications for Industry 4.0 \cite{reqs}, and set the latency threshold to $L_{\text{max}} = 100$ ms. We set the $TTL$ argument of \texttt{local\_aodv+} equal to $2$, we assume a maximum battery capacity of $830$ mAh (3.7 V) and equip the nodes with energy supplies of $E_u^0 = 0-3$ Wh. Last but not least, in order to have a realistic basis for the one-hop latencies $l_{uv}$ to be used in the simulations, we used one-hop propagation measurements achieved by a testbed of WSN430 nodes with CC2420 antenna, under the TinyOS operating system \cite{8390794}. 

In order to benchmark our method, we compared its performance to the performance of the \texttt{PDD} data forwarding method which is provided in \cite{8390794}. %Due to the fact that \texttt{PDD} was designed for static environments without significant network parameter changes, 
 We also compare to a modified version of \texttt{PDD}, which incorporates central reconfiguration when needed (we denote this version as \texttt{PDD-CR}). Specifically, \texttt{PDD-CR} runs \texttt{PDD} until time $t$, when a significant change in the network happens, and then, all network nodes communicate their status ($E^t_u, e_{uv}^t, l_{uv}$) to the network controller $C$ by spending $e_{cc}$ amount of energy. $C$ computes centrally a new (near-optimal as shown in \cite{8390794}) data forwarding plan and the nodes run the new plan. In our case, we run the \texttt{PDD-CR} reconfigurations for each case where we would do the same if we were running \texttt{DistrDataFwd}. As noted before, the conditions that trigger a change of the forwarding paths are either node related (a node dies) or link related (in our case for example, a change of interference which results in significantly increased energy costs, such as $\frac{\epsilon_{uv}^t - \epsilon_{uv}^{t-\tau}}{\epsilon_{uv}^t} > 0.5$)\footnote{The qualitative behavior would not change in case of additional reconfiguration events, which simply increase the number of reconfigurations.}. Finally, we also used IETF RPL \cite{rfc6550} as an additional benchmark. Due to its widespread acceptance in the community as the prevalent version, we used the non-storing version of RPL, in which all the data are routed through the central network controller in our model. This centralized routing manner gives a natural disadvantage of RPL compared to the rest of the methods.

\begin{figure*}[t!]
\centering
    \begin{subfigure}[b]{0.66\columnwidth}
    \centering
        \includegraphics[width=\columnwidth]{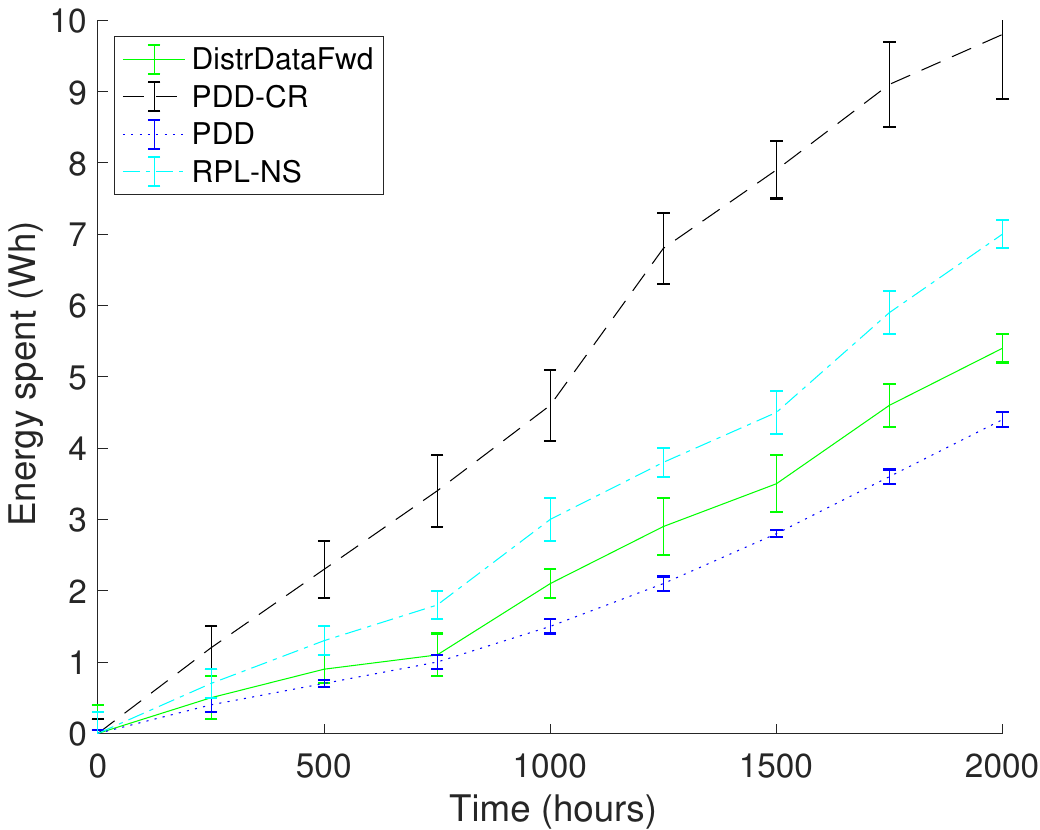}
        \caption{Energy consumption.}
        \label{fig::energy}
    \end{subfigure}
    \begin{subfigure}[b]{0.66\columnwidth}
        \centering
        \includegraphics[width=\columnwidth]{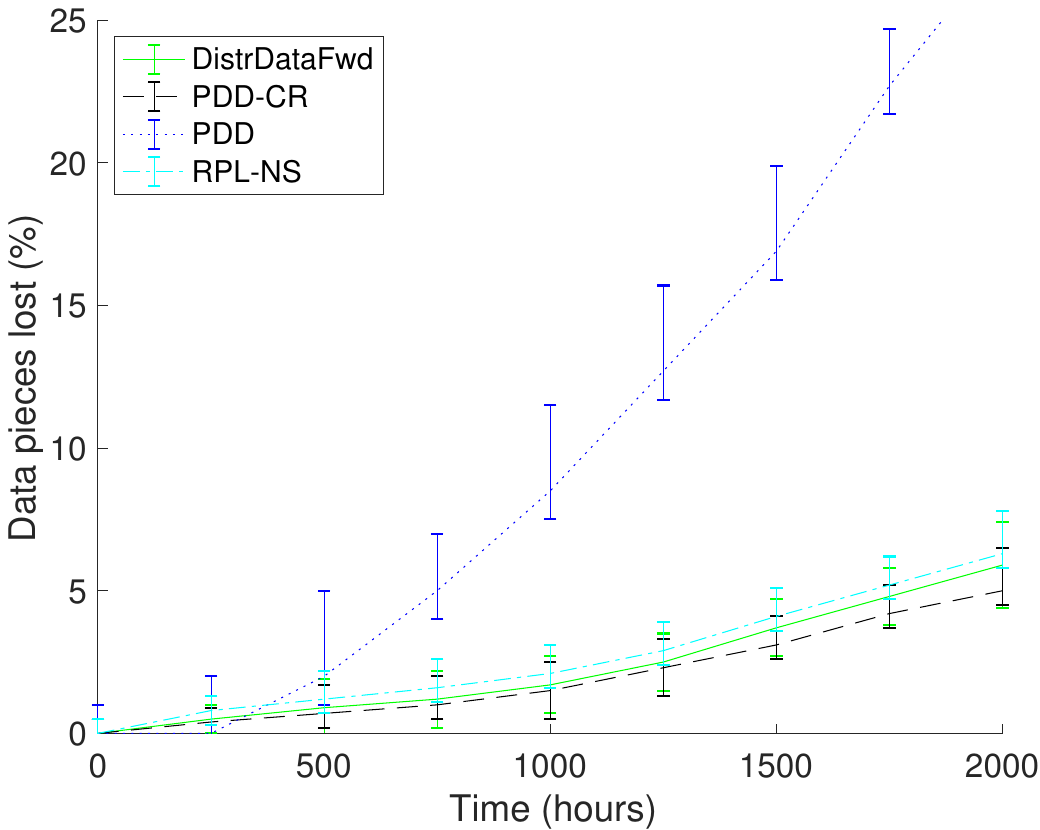}
        \caption{Data pieces lost.}
        \label{fig::data}
    \end{subfigure}
    \begin{subfigure}[b]{0.66\columnwidth}
    \centering
        \includegraphics[width=\columnwidth]{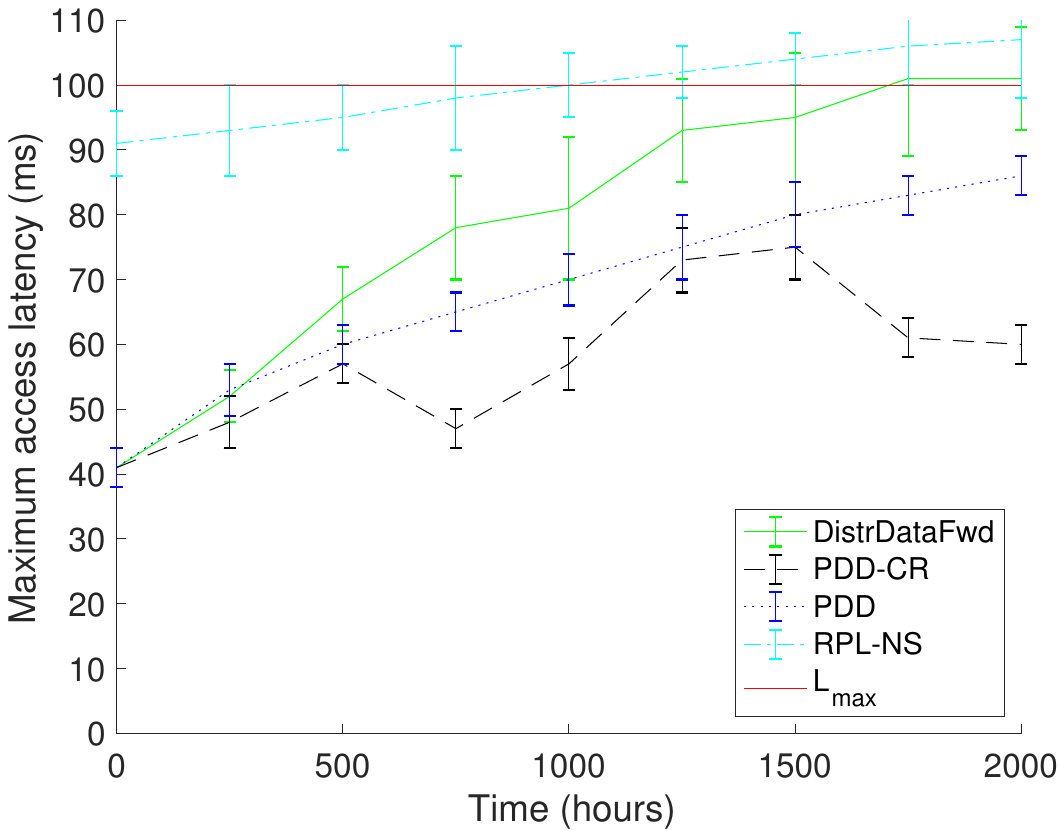}
        \caption{Max. access latency.}
        \label{fig::latency}
    \end{subfigure}
    \caption{Performance results when nodes and links can go offline.}\label{fig::results}
%    \vspace{-0.7cm}
\end{figure*}

\begin{figure*}[t!]
\centering
    \begin{subfigure}[b]{0.66\columnwidth}
    \centering
        \includegraphics[width=\columnwidth]{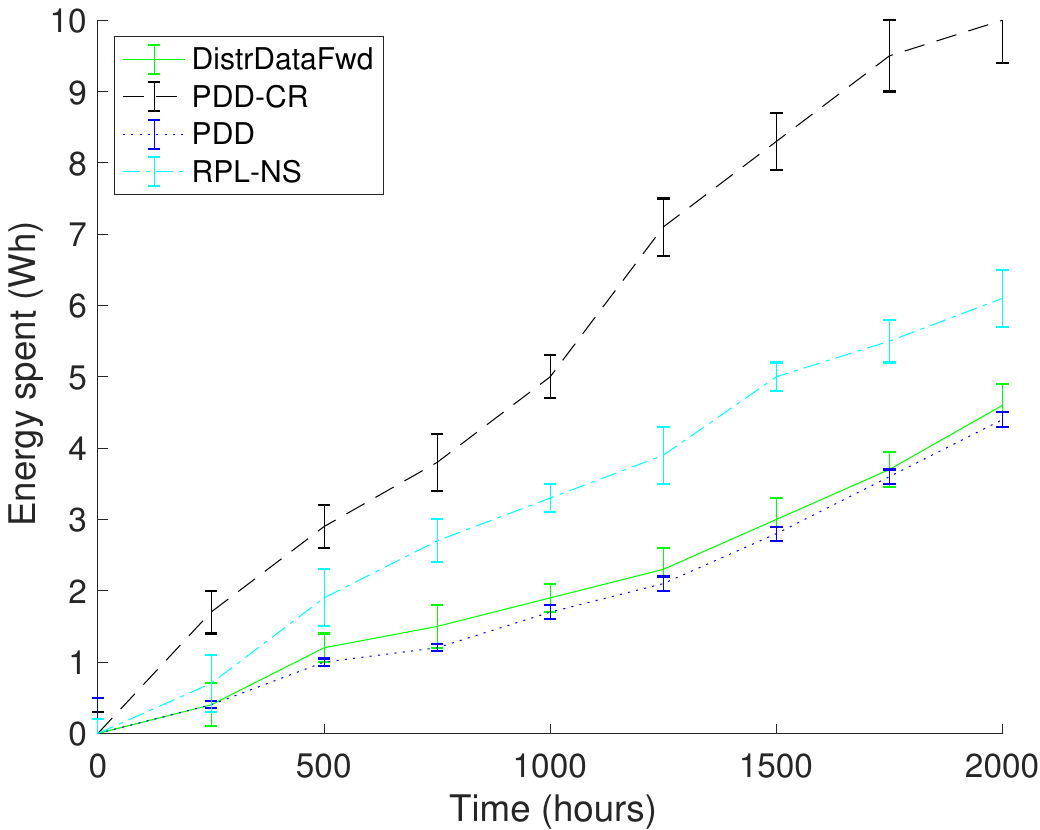}
        \caption{Energy consumption.}
        \label{fig2::energy}
    \end{subfigure}
    \begin{subfigure}[b]{0.66\columnwidth}
        \centering
        \includegraphics[width=\columnwidth]{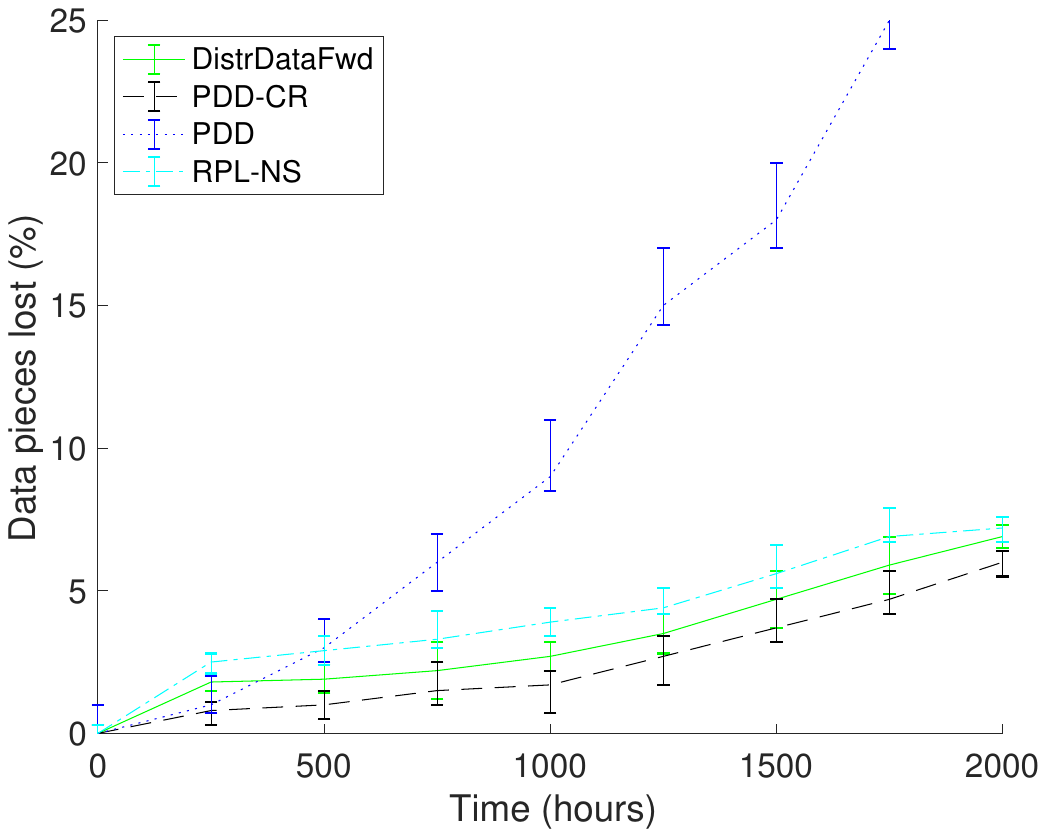}
        \caption{Data pieces lost.}
        \label{fig2::data}
    \end{subfigure}
    \begin{subfigure}[b]{0.66\columnwidth}
    \centering
        \includegraphics[width=\columnwidth]{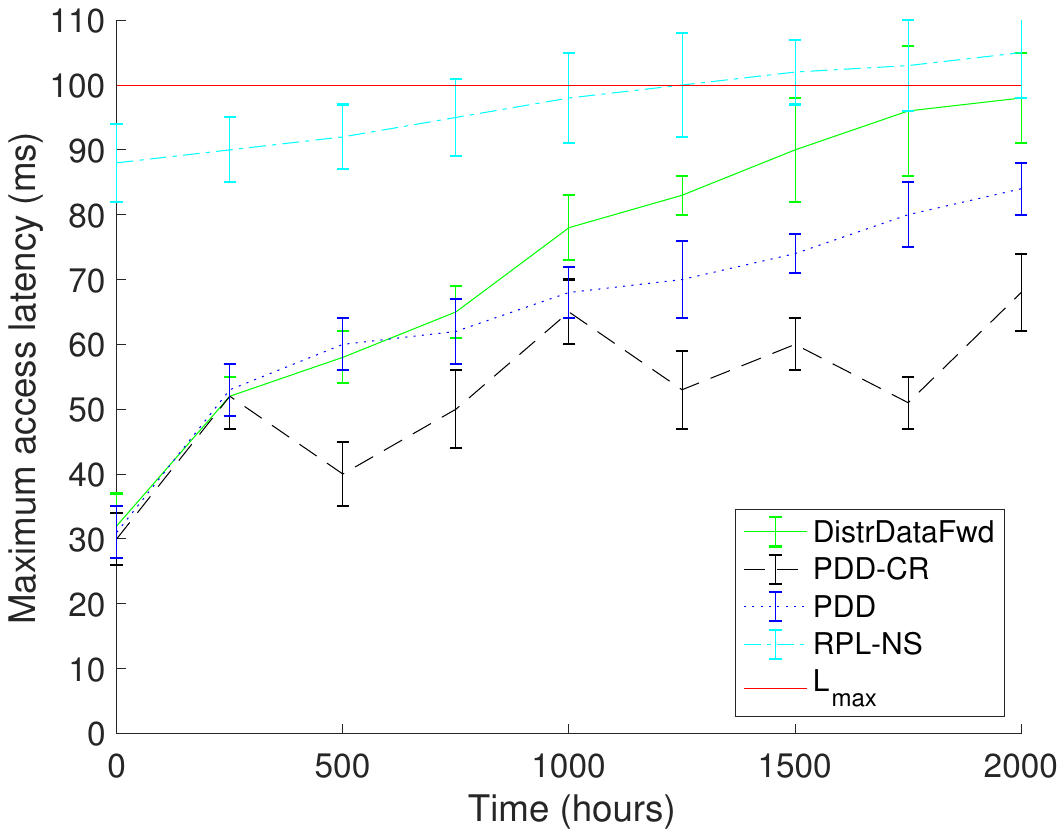}
        \caption{Max. access latency.}
        \label{fig2::latency}
    \end{subfigure}
    \caption{Performance results when nodes can return online.}\label{fig2::results}
%    \vspace{-0.7cm}
\end{figure*}

We ran the simulations for 50 times and we depict the average values, as well as the confidence intervals, so as to capture stochastic variations coming from phenomena such as the variation on the one-hop latencies or the variable path lengths. Specifically, we conduct and discuss two sets of simulations. In the first case (shown in Figure~\ref{fig::results}), we simulate only node and link disactivations, without the possibility of a node to return back online. In this case, the algorithmic functions presented in section \ref{sec::off} are used over time. In the second case (shown in Figure~\ref{fig2::results}), we introduce also the possibility of nodes returning back to an operational state, after they have been disactivated. In this case, the algorithmic functions presented in both section \ref{sec::off} and section \ref{sec::return} are used over time. Note that, in the second case, we initiate the simulations with the same network topology and parameters as in the first case, but with some nodes already disactivated. For this reason, in the beginning of the executions, there is a lower availability of active nodes in the network, so all algorithms start in a disadvantageous state. The performance is of course variating in time, as nodes go on and off during the execution, but, due to the fact that none of the presented algorithms is an exact optimization algorithm with provable guarantees, the initial lower node availability might have an impact on their performance over time.

\begin{figure}[t!]
    \centering
        \includegraphics[width=0.66\columnwidth]{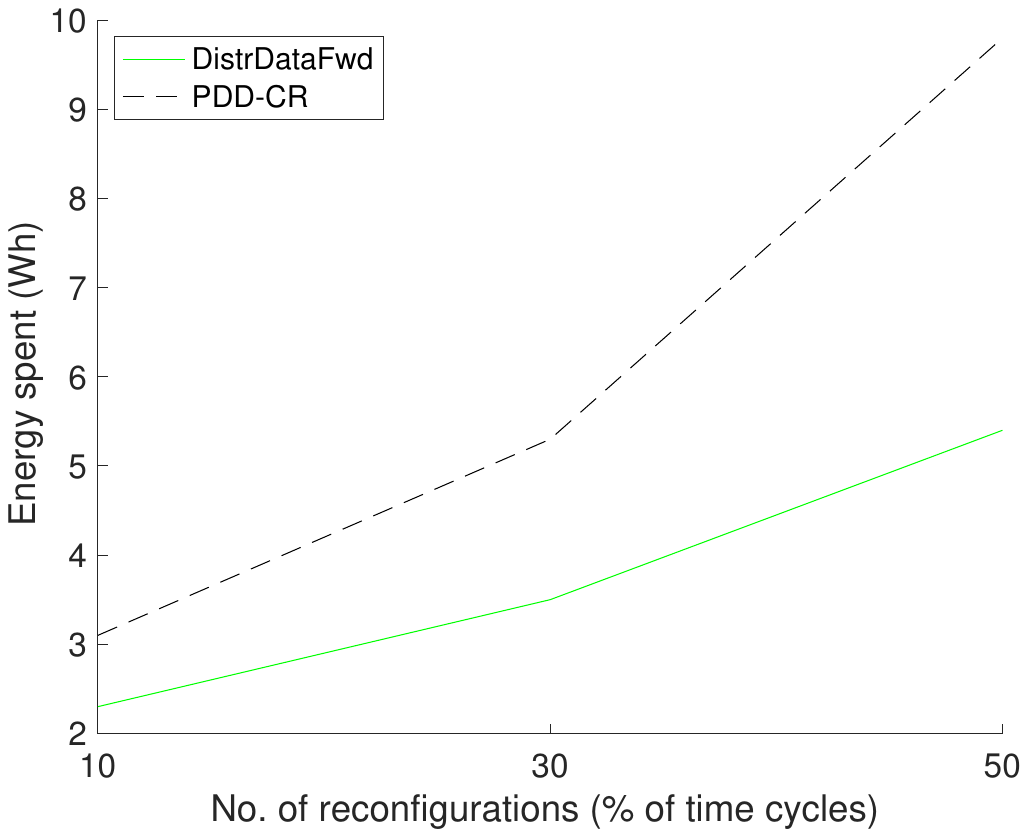}
        \caption{Energy consumption over reconfiguration number.}
        \label{fig::reconfig}
\end{figure}

\textbf{Energy efficiency.} The energy consumption over the entire network during 2000 hours of operation is depicted in Figure~\ref{fig::energy} for the first simulation case and in Figure~\ref{fig2::energy} for the second simulation case. The energy consumption values include the energy consumed for both the data distribution process and the reconfiguration. Our method achieves comparable energy consumption as \texttt{PDD}, despite being a local, adaptive method. This is explained by the following facts: \texttt{PDD-CR} requires more energy than \texttt{DistrDataFwd} for the path reconfiguration process, as during each epoch alteration every node has to spend $e_{cc}$ amount of energy for the configuration phase. On the contrary, in the \texttt{DistrDataFwd} case, only some of the nodes have to participate in a new configuration phase (usually the nodes in the neighborhood of the problematic node), and spend significantly less amounts of energy. In the case of \texttt{PDD}, the nodes do not participate in configuration phases, so they save high amounts of energy. Then main performance difference between Figure~\ref{fig::energy} and Figure~\ref{fig2::energy}, is that the performance of \texttt{DistrDataFwd} approximates better the performance of \texttt{PDD}. This is natural, since the capability of nodes to return back to operational state, and the criterion of replacing nodes in the paths according to the lifetime-based metric (Algorithm \ref{algo::Revive}, line \ref{revive::lifetime}), leads to a better energy consumption across the network. Another interesting conclusion is that greater numbers of reconfigurations (when nodes are returning in the network), lead to more configuration energy needed over time. In fact, in Figure~\ref{fig::reconfig}, we can also see the energy consumption of \texttt{DistrDataFwd} and \texttt{PDD-CR} for different percentages of reconfigurations (w.r.t. the number of time intervals). It is clear that the more the reconfigurations that we have in the network, the more the gap between the performance of \texttt{DistrDataFwd} and \texttt{PDD-CR} increases. Finally, we can see that \texttt{DistrDataFwd} outperforms also RPL. This is natural, since RPL follows longer routing paths through the network controller, and consequently necessitates additional energy resources over the network, an issue that has been discussed also in previous works like \cite{6841029} and \cite{ZhaoKCL17}.

\textbf{Data delivery.} The data pieces lost during 2000 hours of operation are depicted in Figure~\ref{fig::data}. We consider a data piece as lost when the required nodes or path segments are not being available anymore so as to achieve a proper delivery. We can see that the low energy consumption of the \texttt{PDD} method comes at a high cost: it achieves a significantly lower data delivery rate than the \texttt{PDD-CR} and the \texttt{DistrDataFwd} methods. This is natural, because as noted before, \texttt{PDD} computes an initial centralized paths configuration and follows it throughout the entire data distribution process. The performance of the \texttt{DistrDataFwd} method stays very close to the performance of the \texttt{PDD-CR} and RPL methods, which demonstrates the efficiency of \texttt{DistrDataFwd} in terms of successfully delivering the data pieces. The performance differences between Figure~\ref{fig::data} and Figure~\ref{fig2::data}, are not very high among the various data distribution solutions. A difference between the two figures is a slightly increased data piece loss rate for all the solutions in the second case of simulations. As mentioned earlier, this conclusion can be attributed to the fact that the initial availability of nodes in the second case is lower, due to the fact that some of the nodes are already disactivated in the beginning of the executions.

\textbf{Maximum data access latency.} The maximum data access latency during 2000 hours of operation is depicted in Figure~\ref{fig::latency} for the first simulation case and in Figure~\ref{fig2::latency} for the second simulation case. The measured value is the maximum value observed across all the consumers which are part of a connected data delivery path. A particularly important observation is the time when the first pair of producer and consumer exceeds the latency threshold (naturally this happens with \texttt{DistrDataFwd} which do not provide latency preservation guarantees). In the case of \texttt{PDD}, we display the maximum latency reported among all current connected paths as well. Due to the fact that  \texttt{PDD} does not provide a reconfiguration mechanism, when a path gets disconnected, the data pieces are not delivered anymore and its data delivery latency is virtually infinite. Therefore, \texttt{PDD} does not perform well, due to the fact that it is prone to early disconnections without reconfiguration functionality, but as shown in the Figures, for the paths that it manages to keep connected, the latency is kept below the latency threshold. This important weakness of \texttt{PDD} is reflected in Figures \ref{fig::data} and \ref{fig2::data}, in which \texttt{PDD} is significantly outperformed by \texttt{DistrDataFwd} and \texttt{PDD-CR}, especially after the 250 hour mark (given the settings of Table \ref{tab::parameters}, the first path disconnection for \texttt{PDD} is usually happening between 250 and 500 hours). The fluctuation of \texttt{PDD-CR}'s curve is explained by the re-computation from scratch of the data forwarding paths which might result in entirely new data distribution patterns in the network. \texttt{DistrDataFwd} respects the $L_{\text{max}}$ threshold for most of the time. However, in the first simulation case, after 1700 hours of network operation the first producer-consumer pair that exceeds the threshold is reported in most of the simulation executions (based on the settings of Table \ref{tab::parameters}). The pair is not always the same (as the executions are randomized), but as soon as there is a pair violating the threshold, it is enough for the plot line of \texttt{DistrDataFwd} in Figure \ref{fig::latency} to cross over the red line. On the contrary, \texttt{PDD-CR} does not exceed the threshold in this case. This performance is explained by the fact that \texttt{DistrDataFwd}, although efficient, does not provide any strict guarantee for respecting $L_{\text{max}}$, for all producer-consumer pairs, mainly due to the absence of global knowledge on the network parameters during the local computations. \texttt{PDD-CR}, with the expense of additional energy for communication, is able to centrally compute near optimal paths and consequently achieve the desired latency. Interestingly enough, in the second simulation case, \texttt{DistrDataFwd} is not exceeding the threshold (because of returning nodes which ``correct'' the previous latencies) and \texttt{PDD-CR}'s curve is even more variable than than in the first case (once again, because of the returning nodes which provide additional reconfiguration alternatives). \texttt{DistrDataFwd} outperforms RPL in terms of maximum latency observed in the network. This happens because of two reasons: at first, RPL does not receive as input any latency requirements, so the respect of related constraints is not included in the design of the protocol. Secondly, the routing through the network controller increases the number of hops until the data delivery, increasing at the same time the end-to-end latency.

\section{Conclusion} \label{sec::conc}

We identified the need for a distributed reconfiguration method for data forwarding paths in industrial networks. Given the operational parameters of the network, we provided several efficient algorithmic functions which reconfigure the paths of the data distribution process, when a network node fails, and when a network node returns online. The functions regulate how the local path reconfiguration is implemented, ensuring that there will be no loops. We demonstrated the performance gains of our method in terms of energy consumption and data delivery success rate compared to other state of the art solutions. There are two simple ways of improving \texttt{DistrDataFwd}'s performance in terms of respecting the $L_{\text{max}}$ threshold: 
\begin{itemize}
\item insert strict latency checking mechanisms in the \texttt{local\_aodv+} function, with the risk of not finding appropriate (in terms of latency) path replacements, and thus lowering the data delivery ratio due to disconnected paths
\item increase the $TTL$ argument of \texttt{local\_aodv+}, with the risk of circulating excessive amounts of route discovery messages, and thus increasing the energy consumption in the network. 
\end{itemize}
The design and simulative evaluation of those mechanisms can be an interesting topic for future work. Additionally, there are some open issues regarding more accurate applicability of our methods in application oriented industrial field deployments:
\begin{itemize}
\item Due to the fact that we assume a perfect link reliability scenario, we include no retransmission scheduling in the design of the \texttt{ModifyPath} algorithm. However, the abstraction of the wireless network as a graph has been widely discussed in the literature; mainly as the links between the nodes can be quite dynamic, especially in especially in the case of (not so industrial shopfloor tailored) ad hoc lossy low-power networks. This different link reliability assumption necessitates the design of alternative methods for keeping the paths loop free, focusing on efficient retransmission scheduling.
\item In the case where a protocol designer wishes to align our methods to a specific industrial wireless standard, additional design elements should be taken care of. For example, in the case of WirelessHART, the channel scheduling mechanism has to be adapted according to the application requirements. This is because in WirelessHART the channel scheduling mechanism has yet to be standardized, and just some requirements are indicated in the standard document \cite{Nobre_2015}. So, two indicative design considerations are how to assign slots to pairs of nodes in order to avoid collisions, and how to form the WIrelessHART superframe \cite{8793020}. Those considerations could be dealt with methods like graph coloring algorithms on the tree topology, with different constraints to be fulfilled and slot assignment in the superframe with the aim of maximizing the sleep time of nodes and minimizing latency.
\end{itemize}

%\subsubsection*{Acknowledgments.}

%This work was funded by the European Commission through the FoF-RIA Project \emph{AUTOWARE: Wireless Autonomous, Reliable and Resilient Production Operation Architecture for Cognitive Manufacturing} (No. 723909).

%
% ---- Bibliography ----
%

\section*{Acknowledgments} 

This work was funded by the EC through the FoF-RIA Project AUTOWARE (No. 723909).

%\section*{References}

%\bibliographystyle{elsarticle-num}
%\bibliography{refs}{}

\end{document}